\documentclass[acmlarge]{acmart}

\renewcommand\footnotetextcopyrightpermission[1]{} 

\AtBeginDocument{%
  \providecommand\BibTeX{{%
    \normalfont B\kern-0.5em{\scshape i\kern-0.25em b}\kern-0.8em\TeX}}}

\setcopyright{acmcopyright}
\copyrightyear{2022}
\acmYear{2022}

\acmJournal{JOCCH}
\acmVolume{-}
\acmNumber{-}

\usepackage{graphicx}
\usepackage{url}
\usepackage{color}
\usepackage{hyperref}
\usepackage{booktabs} 
\usepackage{makecell}
\usepackage{paralist}
\usepackage{fancyvrb}
\usepackage{lipsum}
\usepackage{listings}

\newcommand{\q}[1]{\lq\lq{}{}#1\rq\rq{}{}}
\newcommand{\sq}[1]{\lq{}#1\rq{}}

\begin{document}

\title[The SeaLiT Ontology]{The SeaLiT Ontology -- An Extension of CIDOC-CRM for the Modeling and Integration of Maritime History Information}

\author{Pavlos Fafalios}
\email{fafalios@ics.forth.gr}
\orcid{0000-0003-2788-526X}

\author{Athina Kritsotaki}
\email{athinak@ics.forth.gr}
\orcid{0000-0001-5086-7284}

\author{Martin Doerr}
\email{martin@ics.forth.gr}
\orcid{0000-0002-6006-0844}

\affiliation{%
  \institution{Centre for Cultural Informatics and Information Systems Laboratory, FORTH-ICS}
  \streetaddress{N. Plastira 100}
  \city{Heraklion}
  \country{Greece}
  \postcode{GR-70013}
}

\begin{abstract}
We describe the construction and use of the SeaLiT Ontology, an extension of the ISO standard CIDOC-CRM for the modelling and integration of maritime history information. 
The ontology has been developed gradually, following a bottom-up approach that required the analysis of large amounts of real primary data (archival material) as well as knowledge and validation by domain experts (maritime historians).  
We present the specification of the ontology, RDFS and OWL implementations, as well as knowledge graphs that make use of this data model for integrating information originating from a large and diverse set of archival documents, such as crew lists, sailors registers, naval ship registers and payrolls. We also describe an application that operates over these knowledge graphs and which supports historians in exploring and quantitatively analysing the integrated data through a user-friendly interface. Finally, we discuss aspects related to the use, evolution and sustainability of the ontology. 
\end{abstract}



\maketitle

\section{Introduction}

Maritime history is the study of human activity at sea. It covers a broad thematic element of history, focusing on understanding humankind's various relationships to the oceans, seas, and major waterways of the globe \cite{hattendorf2012maritime}.
A large area of research in this field requires the collection and integration of data coming from multiple and diverse historical sources, in order to perform qualitative and quantitative analysis of empirical facts and draw conclusions on possible impact factors \cite{fafalios2021FastCat,petrakis2021}.

Consider, for instance, the real use case of the SeaLiT project (ERC Starting Grant in the field of maritime history)\footnote{\url{https://sealitproject.eu/}}, which studies the transition from sail to steam navigation and its effects on seafaring populations in the Mediterranean and the Black Sea between the 1850s and the 1920s \cite{delis2020seafaring}. Historians in this project have collected a large number of archival documents of different types and languages, including crew lists, payrolls, sailor registers, naval ship register lists, and employment records, gathered from multiple authorities in different countries (more about this project in Sect.~\ref{subsec:sealit}). 
Complementary information about the same entity of interest, such as a ship, a port, or a captain, may exist in different archival documents. For example, for the same ship, one source may provide information about its owners, another source may provide construction details and characteristics of the ship (length, width, tonnage, horsepower, etc.), while other sources may provide information about the ship's voyages and crew.

Information integration is crucial in this context for performing valid data analysis and drawing safe conclusions, such as finding answers to questions that require combining and aggregating information, like \textit{\q{finding the number of sailors per residence location that arrived at a specific port and who were crew members in ships of a specific type, e.g. Brig}}. 
Moreover, information integration under a common data model can produce data of high value and long-term validity that can be reused beyond a particular research activity or project, as well as integrated with other datasets by the wider (historical science) community.

To this end, this paper describes the construction and use of the \textit{SeaLiT Ontology}. The ontology aims at facilitating a shared understanding of maritime history information by providing a common and extensible semantic framework for information modeling and integration.
It uses and extends the CIDOC Conceptual Reference Model (CRM) (ISO 21127:2014)\footnote{\url{https://cidoc-crm.org/}} as a formal ontology of human activity, things and events happening in space and time \cite{doerr2003cidoc}. 

The ontology was designed considering requirements and knowledge of domain experts (a large group of maritime historians), expressed through research needs, inference processes they follow, and exceptions they make. It was developed in a bottom-up manner by analysing large and heterogeneous amounts of primary data, in particular archival documents of different types and languages gathered from authorities in several countries, including crew lists, payrolls, civil registers, sailor registers, naval ship registers, employments records, censuses, and others. 
All modeling decisions were validated by the domain experts and, in practice, by transforming their data (transcripts) to a rich semantic network based on the SeaLiT Ontology, which enables them (through a user-friendly interface) to find answers to information needs that require combining information of different sources. 

We describe the  methodology and the steps we followed for designing the ontology, and provide its specification, RDFS and OWL implementations, as well as knowledge graphs that make use of the ontology for integrating data transcribed from a large and diverse set of archival documents. We also describe a data exploration application that operates over these knowledge graphs and which currently supports maritime historians in exploring and analysing the integrated data. 

Table \ref{tab:links} provides the key access links to the SeaLiT Ontology as well as related resources and information.

\begin{table}[h]
\begin{center} 
\caption{Key access links and information of the SeaLiT Ontology.}
\label{tab:links}
\vspace{-2mm}
\begin{tabular}{ll}
\toprule
SeaLiT Ontology Specification & \url{https://zenodo.org/record/6797750} \\
DOI of the SeaLiT Ontology & 10.5281/zenodo.6797750\\
Namespace of the SeaLiT Ontology  & \url{http://www.sealitproject.eu/ontology/} \\
SeaLiT Ontology RDFS (Turtle)  & \url{https://sealitproject.eu/ontology/SeaLiT_Ontology_v1.1_RDFS.ttl} \\
SeaLiT Ontology RDFS (RDF/XML)  & \url{https://sealitproject.eu/ontology/SeaLiT_Ontology_v1.1_RDFS.rdf} \\
SeaLiT Ontology OWL (RDF/XML)  & \url{https://sealitproject.eu/ontology/SeaLiT_Ontology_v1.1.owl} \\
\midrule
SeaLiT Knowledge Graphs (KGs)  & \url{https://zenodo.org/record/6460841} \\
DOI of SeaLiT KGs & 10.5281/zenodo.6460841 \\
ResearchSpace application over the KGs & \url{http://rs.sealitproject.eu/} \\  
\midrule
License of SeaLiT Ontology \& KGs & Creative Commons Attribution 4.0 \\ 
\bottomrule
\end{tabular}
\end{center}
\end{table}

The rest of this paper is organised as follows: 
Section~\ref{sec:background} describes the context of this work, provides the required background, and discusses related work. 
Section~\ref{sec:methodology} details the methodology and principles we have followed for building the ontology. 
Section~\ref{sec:ontology} presents the ontology, describes an example on how a part of the model was revised several times to incorporate new historical knowledge, and provides its specification as well as an RDFS and an OWL implementation. 
Section~\ref{sec:application} describes the application of the ontology in a real context.
Section~\ref{sec:usage} discusses its usage and sustainability. 
Finally, Section~\ref{sec:conclusion} concludes the paper and outlines future work.

\section{Context, Background and Related Work}
\label{sec:background}

\subsection{The SeaLiT Project}
\label{subsec:sealit}

The ontology has been developed in the context of the SeaLiT project\footnote{\url{https://sealitproject.eu/}},  a European  project in the field of maritime history (ERC Starting Grant, No 714437). The project studies the transition from sail to steam navigation and its effects on seafaring populations in the Mediterranean and the Black Sea between the 1850s and the 1920s. Historians in SeaLiT investigate the maritime labour market, the evolving relations among ship-owners, captain, crew, and local societies, and the development of new business strategies, trade routes, and navigation patterns, during the transitional period from sail to steam. 
The main concepts on which the scientific research focuses, are the ships (including various information such as type, usage, dimensions, technology), the people related to the ships (sailors, ship owners, students, relatives) and the historical events/activities related to these (such as voyages, recruitments, payments).

The archival sources considered and studied in SeaLiT range from hand written ship log books, crew lists, payrolls and employment records, to registers of different types such as civil, sailors, students and naval ship registers. These archival sources have been gathered from different authorities in countries of the Mediterranean and the Black Sea, and are written in different languages, including Spanish, Italian, French, Russian, and Greek. 
The full archival corpus studied in SeaLiT is described in the project's web site.\footnote{\url{https://sealitproject.eu/archival-corpus}}

\subsection{The ISO standard CIDOC-CRM}
The SeaLiT Ontology uses and extends the CIDOC-CRM (Conceptual Reference Model)\footnote{\url{http://www.cidoc-crm.org/}}, in particular its stable version 7.1.1, which  means that each class of the SeaLiT Ontology is a direct subclass or a descendant of a CIDOC-CRM class. 

CIDOC-CRM is a high-level, event-centric ontology of human activity, things and events happening in spacetime, providing definitions and a formal structure for describing the implicit and explicit concepts and relationships used in cultural heritage documentation \cite{doerr2003cidoc}. 
It is the international standard (ISO 21127:2014)\footnote{\url{https://www.iso.org/standard/57832.html}} for the controlled exchange of cultural heritage information, intended to be used as a common language for domain experts and implementers to formulate requirements for information systems, providing a way to integrate cultural heritage information of different sources. 

\begin{figure}[h]
    \centering
    \fbox{\includegraphics[width=0.8\textwidth]{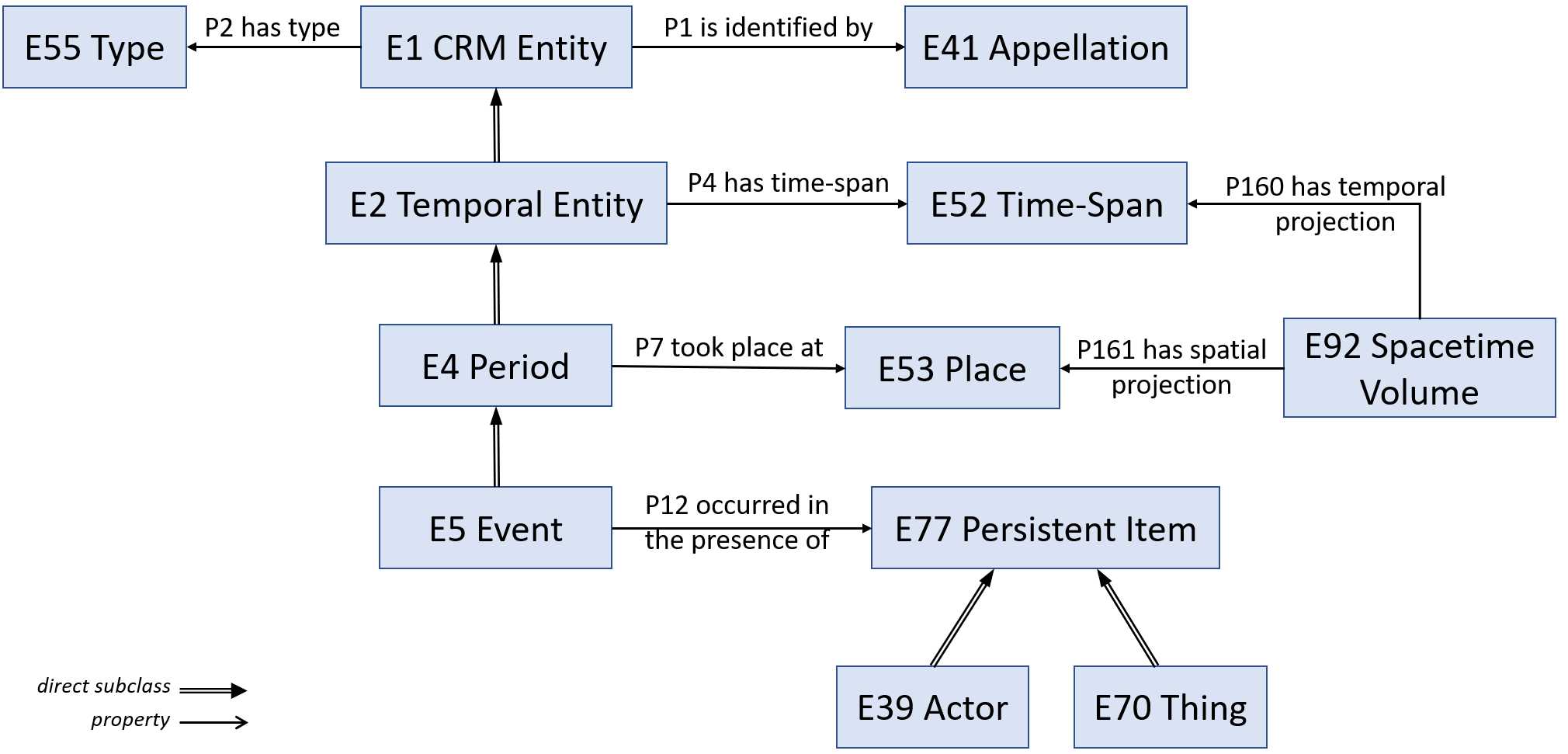}}
    \vspace{-2mm}
    \caption{High level properties and classes of CIDOC-CRM.}
    \label{fig:crm1}
 \end{figure}

The considered stable release of CIDOC-CRM (version 7.1.1) consists of 81 classes and 160 unique properties. The highest-level distinction in CIDOC-CRM is represented by the top-level concepts of {\tt E77 Persistent Item} (equivalent to the philosophical notion of endurant), {\tt E2 Temporal Entity} (equivalent to the philosophical notion of perdurant) and, further, the concept of {\tt E92 Spacetime Volume} which describes the entities whose substance has or is an identifiable, confined geometrical extent in the material world that may vary over time.
Fig.~\ref{fig:crm1} depicts how the high level  classes of CIDOC-CRM are connected. 

\subsection{Related Work} 

Over the last years, methods and technologies of the Semantic Web have started playing a significant and ever increasing role in historical research. 
The survey in \cite{merono2015semantic} reviews the state of the art in the application of semantic technologies to historical research, in particular works related to i) knowledge modeling (ontologies, data linking), ii) text processing and mining, iii) search and retrieval, and iv) semantic interoperability (data integration, classification systems). 

As regards ontologies for the modeling of \textit{maritime history} information, the most relevant work is an ongoing project on the ontology management environment OntoME~\cite{beretta2021challenge} that aims to provide a data model for the field of maritime/nautical history.\footnote{\url{https://ontome.net/namespace/66}} The project is a cooperation between the Huygens Institute for the History of the Netherlands, LARHRA and the Data for History consortium.
The current (draft) model consists of 13 classes and 12 properties, while it makes use of CIDOC-CRM as well as extensions of CIDOC-CRM. The ontology is unfinished and not for use yet (as of December 15, 2022).  

\textit{Conflict}\footnote{\url{http://ontologies.michelepasin.org/docs/conflict/index.html}} is an ontology developed in the context of the SAILS project (2010-2013)\footnote{\url{http://sailsproject.cerch.kcl.ac.uk/}} that models concepts useful for describing the First World War.
The provided ontology version (0.1) is actually a \textit{taxonomy} consisting of 175 classes, some of which allow modeling information related to maritime history, like the classes {\tt Ship}, {\tt Ship\_journey}, {\tt Ship\_type}, and {\tt Ownership}.
Similarly, there are ontologies that could be used for modeling other \textit{parts} of the model, such as \textit{GoodRelations}~\cite{hepp2008goodrelations}, a lightweight ontology for exchanging e-commerce information, for the part that concerns payments for products. 

We selected to use CIDOC-CRM because it is the standard ontology for cultural heritage documentation, extensively used in the fields of cultural heritage, history and archaeology.
It is directly related to the domain of discourse of history, as a discipline that studies the life of humans and societies in the past. This scope, studied from the point of view of maritime historical research, can be represented by the abstraction of reality offered by CIDOC-CRM. 
As an example, we can directly take advantage of the (direct or inherited) properties of the CIDOC-CRM class {\tt E7 Activity}, such as \textit{\sq{P14 carried out by}}, \textit{\sq{P4 has time-span}}, \textit{\sq{P7 took place at}}, etc., and use them for describing instances of classes of the SeaLiT Ontology that are subclasses of {\tt E7 Activity} (e.g. {\tt Voyage}, {\tt Arrival}, {\tt Recruitment}, etc.).
Therefore, using CIDOC-CRM facilitates data integration with relevant (existing or future) datasets that also make use of CIDOC-CRM, but also it enables data sustainability because CIDOC-CRM is a living standard and has a very active community that constantly works on it and improves it.

Finally, there is a plethora of ontologies which have been developed as extensions of CIDOC-CRM, e.g. 
CRMas~\cite{niccolucci2017documenting} for documenting archaeological science, 
CRMgeo~\cite{hiebel2017crmgeo} for geospatial information,  
CRMdig~\cite{theodoridou2010modeling} for provenance of digital objects, 
IAM~\cite{doerr2011factual} for factual argumentation,
and others.

\section{Design Methodology and Principles}
\label{sec:methodology}

\subsection{Overall Methodology}
The ontology has been created gradually, following a bottom-up strategy \cite{gandon2002distributed}, working with real empirical data and information needs, in particular digitised historical records (transcripts) and corresponding data structures in various forms, as well as research questions provided by a large group of historians. 
The archival material together with the research questions define the modeling requirements. 

The main characteristics of our strategy are summarised as follows: 

\begin{itemize}
    \item Study and analysis of a large and diverse set of archival sources related to maritime history. This material provides historical information about ships, persons (such as sailors, captains, ship owners, students), and relevant activities and events (such as voyages, recruitments, payments, teaching activities).
    \item Gathering of research questions and corresponding information needs (\textit{competency questions}) for which the considered archival sources can provide answers or important relevant information. 
    \item Lengthy discussions with a large group of maritime historians from different institutions and countries (Spain, Italy, France, Croatia, Greece), for consulting as well as understanding of inference processes and exceptions they make. 
\end{itemize}

\begin{table}
\begin{center} 
\caption{Considered archival sources and type of recorded information.}
\label{tab:archSources}
\scriptsize
\begin{tabular}{p{3.9cm}|p{10cm}}
\toprule
Archival source & Overview of recorded information and example transcript\\
\midrule
Crew and displacement list (Roll) 
&  ships (name, type, construction location, construction year, registry location, owners), ports of provenance, arrival ports, destination ports, crew members (name, father's name, birth place, residence location, profession, age), embarkation ports, discharge ports. 
\textbf{[example transcript: \url{https://tinyurl.com/4ukzezfe}]} 
\\ \midrule
Crew List (Ruoli di Equipaggio) 
& ships (name, type, construction location, construction year, registry number, registry port, owners), voyages (date from/to, duration, total crew number),  destinations, departure ports, arrival ports, crew members (name, residence location, birth year, serial number, profession), embarkation ports, discharge ports. 
\textbf{[example transcript: \url{https://tinyurl.com/2u35frya}]} 
\\ \midrule
General Spanish Crew List & ships (name, type, tonnage, registry port), ship owners, crew members (name, age, residence location), voyages (date from/to, total crew number), embarkation ports, destinations.
\textbf{[example transcript: \url{https://tinyurl.com/3axs6ret}]}
\\ \midrule
Sailors Register (Libro de registro de marineros) & seafarers (name, father's name, mother's name, birth date, birth place, profession, military service organisation locations) 
\textbf{[example transcript: \url{https://tinyurl.com/2p8kzm6n}]} 
\\ \midrule
Register of Maritime Personnel & persons (name, father's name, mother's name, birth place, birth date, residence location, marital status, previous profession, military service organisation location). 
\textbf{[example transcript: \url{https://tinyurl.com/4v6hnwjj}]} 
\\ \midrule
Seagoing Personnel & persons (name, father's name, marital status, birth date, profession, end of service reason, work status type), ships (name), destinations.
\textbf{[example transcript: \url{https://tinyurl.com/2x5cu37n}]} 
\\ \midrule
Naval Ship Register List & ships (name, type, tonnage, length, construction location, registration location, owner).
\textbf{[example transcript: \url{https://tinyurl.com/bdhx87tr}]} 
\\ \midrule
List of Ships  & ships (name, previous name, type, registry port, registry year, construction place, construction year, tonnage, engine construction place, engine manufacturer, nominal power, indicated power, owners).
\textbf{[example transcript: \url{https://tinyurl.com/2cphfpef}]} 
\\ \midrule
Civil Register & persons (name, profession, origin location, age, sex, marital status, death location, death reason, related persons).
\textbf{[example transcript: \url{https://tinyurl.com/bdzeja8n}]} 
\\ \midrule
Maritime Register, La Ciotat & persons (name, birth date, birth place, residence location, profession, service sector), embarkation locations, disembarkation locations, ships (name, type, navigation type), captains, patrons.
\textbf{[example transcript: \url{https://tinyurl.com/fkhyyp4a}]} 
\\ \midrule
Students Register & students (origin location, profession, employment company, religion, related persons), courses (title, subject, date from/to, semester, total number of students).
\textbf{[example transcript: \url{https://tinyurl.com/mryp6cbb}]} 
\\ \midrule
Census La Ciotat & occupants (name, age, birth year, birth place, nationality, marital status, religion, profession, working organisation, household role, address).
\textbf{[example transcript: \url{https://tinyurl.com/4dzfcbtt}]} 
\\ \midrule
Census of the Russian Empire & occupants (name, patronymic, sex, age, marital status, estate, religion, native language, household role, occupation, address).
\textbf{[example transcript: \url{https://tinyurl.com/43xczvux}]} 
\\ \midrule
Payroll (of Greek Ships) & ships (name, type, owners), captains, voyages (date from/to, total days, days at sea, days at port, overall total wages, overall pension fund, overall net wage), persons (name, adult/child, literacy, origin location, professio/rank), employments (recruitment date, discharge date, recruitement location, monthy wage, total wage, pension fund, net wage).
\textbf{[example transcript: \url{https://tinyurl.com/ztjk4jw7}]} 
\\ \midrule
Payroll (of Russian Steam Navigation and Trading Company) & ships (name, owners), persons (name, patronymic, adult/child, sex, birth date, estate, registration place), recruitments (port, type of document, rank/specialisation, salary per month).
\textbf{[example transcript: \url{https://tinyurl.com/y5urjhc9}]} 
\\ \midrule
Employment records (Shipyards of Messageries Maritimes, La Ciotat) & workers (name, sex, birth year, birth place, residence location, marital status, profession, status of service in company, workshop manager).
\textbf{[example transcript: \url{https://tinyurl.com/yc3havkc}]} 
\\ \midrule
Logbook  & ships (name, type, telegraphic code, tonnage, registry port, owners), captains, departure ports, destination ports, route movements, calendar event types.
\textbf{[example transcript: \url{https://tinyurl.com/mrx2re9k}]} 
\\ \midrule
Accounts Book  & ships (name, type, owners), voyages, captains, departure ports, destination ports, ports of call, transactions (type, recording location, supplier, mediator, receiver). 
\textbf{[example transcript: \url{https://tinyurl.com/4uf3bye8}]} 
\\
\bottomrule
\end{tabular}
\end{center}
\end{table}

In more detail, our approach focused on studying and analysing the historical sources from the historians perspective, following their respective research questions and practices of documentation. 
In order to achieve that, we had to consult all the data providers (coming from different research teams and countries) for a long period and to do extensive research on their research practices and the historical data for the development and the validation of the model. 
As a result, the model was designed from actual data values, from existing (and used) structured information sources (such as spreadsheets) and historical records (transcripts) that include the original information. The model's concepts were refined several times during the span of the project for considering new information coming from new kinds of sources.
Table~\ref{tab:archSources} provides the considered archival sources as well as an overview of the recorded information and an example record (transcript) for each source.\footnote{A web application that allows exploring the data in the transcripts of these archival sources is available at:   \url{https://catalogues.sealitproject.eu/}}

As regards the research questions and information needs provided by the historians, their majority concerns aggregated information, such as \textit{number of sailors per origin location that arrived at a specific port}, \textit{average tonnage of ships}, \textit{wage level per country}, \textit{percentages of immigration in relation to the sailors' profession}, etc. 
Other information needs concern the retrieval of a specific list of entities (e.g. \textit{ship construction places during a specific time period}), comparative information  (e.g. \textit{time of sailors' service in relation to the time on land}, \textit{number of  women/men in ships}, etc.), or the retrieval of a specific value (e.g. \textit{total number of officers employed by the company in a specific year or span of years}).\footnote{The full list of information needs is available at \url{https://users.ics.forth.gr/~fafalios/SeaLiT_Competency_Questions_InfoNeeds.pdf}}

For creating the ontology, we followed a custom engineering methodology ~\cite{kotis2020ontology} which, though, maintains most of the features supported by existing methodologies, such as HCOME~\cite{kotis2006human} and DILIGENT~\cite{pinto2009ontology}. In particular:
\begin{itemize}
    \item Data-driven / bottom-up processing (our strategy for the development of the ontology) 
    \item Involvement of domain experts (maritime historians in our case)
    \item Iterative processing (gradual, highly-iterative ontology development)
    \item Collaborative engineering processing (within a small team of conceptual modeling experts)
    \item Validation and exploitation (validation by domain experts and application in a real context)
    \item Detailed versioning (multiple intermediate versions, currently in stable version 1.1)
\end{itemize}

\subsection{Design Steps and Principles}

The basis for the model was CIDOC-CRM since it is a standard suitable for recording historical information relating who, when, where, and what. 
From an ontological point of view, we followed the below steps:

\begin{enumerate}
    \item We have extended CIDOC-CRM by creating new classes as subclasses of CIDOC-CRM classes and defining properties accordingly (with some of them being subproperties of CIDOC-CRM properties).
    After extending or revising the model for a given type of archival source and corresponding information needs, we created mappings for transforming the data from the source schema to a semantic network (RDF triples) based on the designed (target) model. This conceptual alignment was an important step to the ontology development process, contributing to redesign concepts and finalise the model.

    \item We distinguished the entities included in the existing schemata into those that directly or indirectly imply an \textit{event} and to those that imply \textit{objects}, mobile or immobile, and classified them in abstraction levels according to whether they represent individuals, or set of individuals. We realised that most binary relationships acquire substance as temporal entities (e.g. \textit{has met}, \textit{has created}, etc.). This principle helped us to detect hidden events in the data structures. 
    
    \item We classified the existing relations between the entities according to the abstraction level which their domain and range entity belong to, and created class and property hierarchies accordingly.
    We did not define the same property twice for different classes, but found the most general (super)class that the property applies to. The discovery of repeating properties for different classes, suggested that they rely on a common, more general concept, causal to the ability to have such a relation in the first place. Finding the single most general concept to describe this common generalization allowed the creation of a general class to which the properties can be applied and from which these relations can be inherited by assigning the originally modelled classes as subclasses of the newly created generalization (like in the case of classes {\em Money for Service} and {\em Legal Object Relationship}). 
    
    \item We found classes for the relevant properties, and not properties for relevant classes (e.g. \textit{Voyage} for the property \sq{voyages}, \textit{Ship Construction} for \sq{constructed}, etc.). We detected the general classes for which each property is characteristic of. In other terms, we found the one most specific class that generalizes over all classes for which the property applies as domain or range.
    
    \item We defined concepts by finding the identity criteria of them, by distinguishing what is and what is not an instance of these concepts. We identified classes that exist independent from the property, and not \q{anything that has this property} (e.g. the case of the \textit{Service} concept). 
    
    \item The number of the classes and relationships developed can answer queries of \textit{global} nature. By global queries we mean those that users would address to more than one database (source) at the same time in order to get a comprehensive answer, in particular including joins across databases. 
    It should also be emphasised that the goal was not to model \sq{everything} but rather to model the necessary and well understood concepts  for this specific domain.
    
\end{enumerate}

The ontology was built following these principles. 
Its design and development was an iterative process with several repetitions of the steps described above.


\section{The SeaLiT Ontology}
\label{sec:ontology}

We first provide an overview of the ontology (Sect.~\ref{subsec:ontOverview}), then we describe an ontology evolution example (Sect.~\ref{subsec:evolution}), and finally we present the specification of the ontology as well as RDFS and OWL implementations (Sect.~\ref{subsec:specAndRdfs}).

\subsection{Ontology Overview}
\label{subsec:ontOverview}

The ontology currently (version 1.1) contains 46 classes, 79 properties and 4 properties of properties, allowing the description of information about \textit{ships}, \textit{ship voyages}, \textit{seafaring people}, \textit{employments} and \textit{payments}, \textit{teaching activities}, as well as a plethora of other related activities and characteristics.
Appendices \ref{appendix:A} and \ref{appendix:B} provide the full class and property hierarchy, respectively.

\begin{figure}
    \centering
    \fbox{\includegraphics[width=15cm]{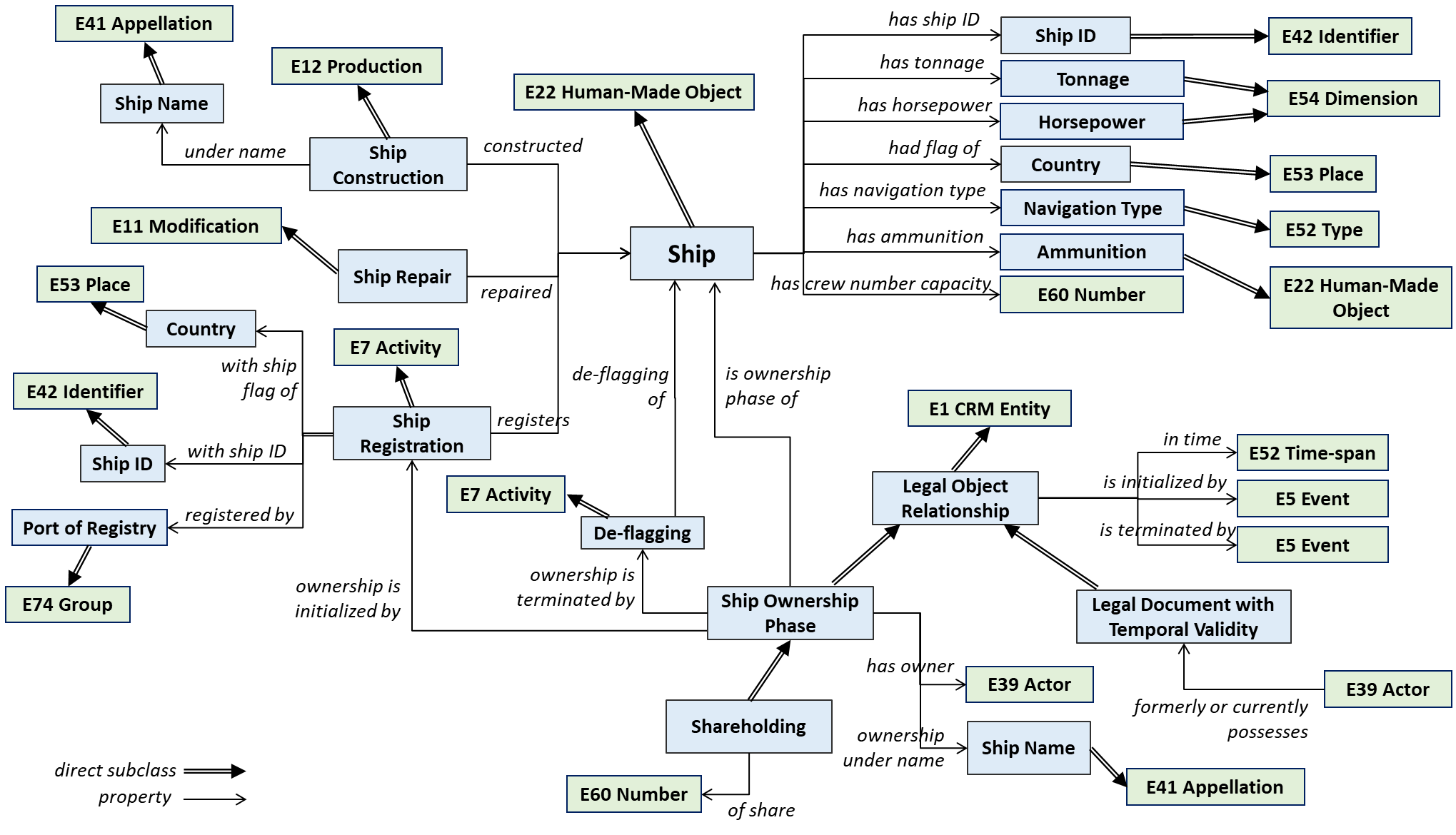}}
    \vspace{-2mm}
    \caption{Modelling information about a ship.}
    \label{fig:model_ship}
\end{figure}

Fig.~\ref{fig:model_ship} shows how information about a \textit{ship} is modelled.\footnote{The classes whose name starts with the letter 'E' followed by a number are CIDOC-CRM classes (these are in green boxes in the figures). All other are classes of the SeaLiT Ontology (in blue boxes). Accordingly, all properties whose name starts with the letter 'P' followed by a number are properties of CIDOC-CRM, while all other are properties of the SeaLiT Ontology.} 
A {\tt Ship} (subclass of {\tt E22 Human-Made Object}) is the result of a {\tt Ship Construction} activity (subclass of {\tt E12 Production}) which gave the {\tt Ship Name} (subclass of {\tt E41 Appellation}) to the ship. A ship also has some characteristics, like {\tt Horsepower} and {\tt Tonnage} (subclasses of {\tt E54 Dimension}; this allows providing, apart from the value, the corresponding measurement unit, a note, etc.), and is registered through a {\tt Ship Registration} (subclass of {\tt E7 Activity}) by a {\tt Port of Registry} (subclass of {\tt E74 Group}), with a ship flag of a particular {\tt Country} (subclass of {\tt E53 Place}) and with a particular {\tt Ship ID} (subclass of {\tt E42 Identifier}).  
Modeling the ship ID as a class allows including additional information about the identifier, such as which authority provided the identifier, when, etc. (by connecting it to the CIDOC-CRM class {\tt E15 Identifier Assignment}).  
Finally, a ship has one or more {\tt Ship Ownership Phase}s (subclass of {\tt Legal Object Relationship}), each one initialized by a {\tt Ship Registration} and terminated by a {\tt De-flagging} activity.  Note here that, all classes related to activities (like {\tt Ship Construction}, {\tt Ship Repair}, {\tt De-flagging}, etc.) can make use of the CIDOC-CRM property {\em \sq{P4 has time-span}} for providing temporal information.

\begin{figure}
    \centering
    \fbox{\includegraphics[width=14cm]{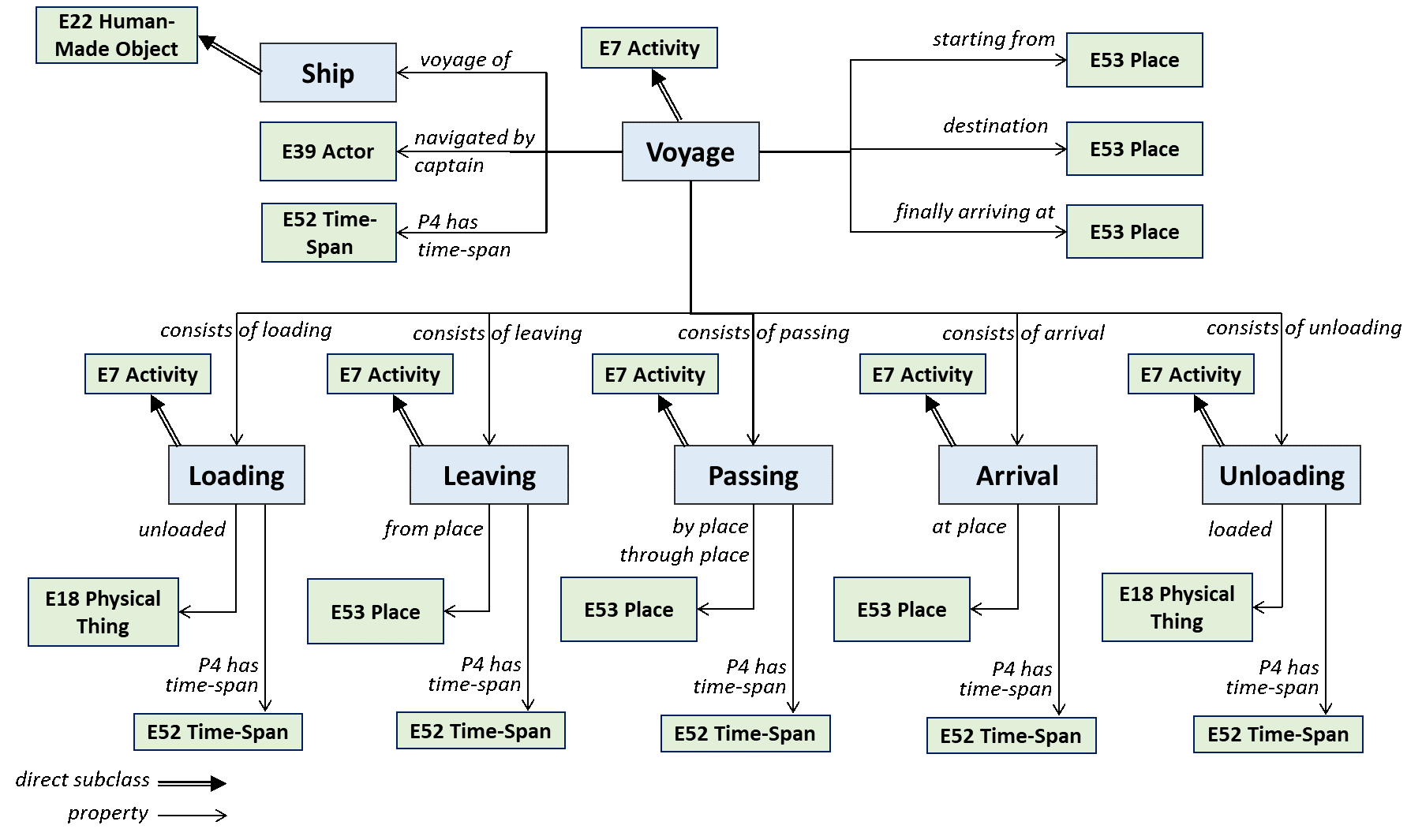}}
    \vspace{-2mm}
    \caption{Modelling information about a ship voyage.}
    \label{fig:model_voyage}
\end{figure}

Fig.~\ref{fig:model_voyage} shows how information about a \textit{ship voyage} is modelled in the ontology. First, a {\tt Voyage} (subclass of {\tt E7 Activity}) concerns a particular Ship, navigated by one or more captains ({\tt E39 Actor}), and has a \textit{starting from} place, a \textit{destination} place, and a \textit{finally arriving at} place ({\tt E53 Place}).  Then, the main activities during a ship voyage include {\tt Loading} things, {\tt Leaving} from a place, {\tt Passing} by or through a place, {\tt Arrival} at a place, and {\tt Unloading} things. All these activities are  linked to a {\tt E52 Time-Span} through the CIDOC-CRM property {\em \sq{P4 has time-span}}.

\begin{figure}
    \centering
    \fbox{\includegraphics[width=15cm]{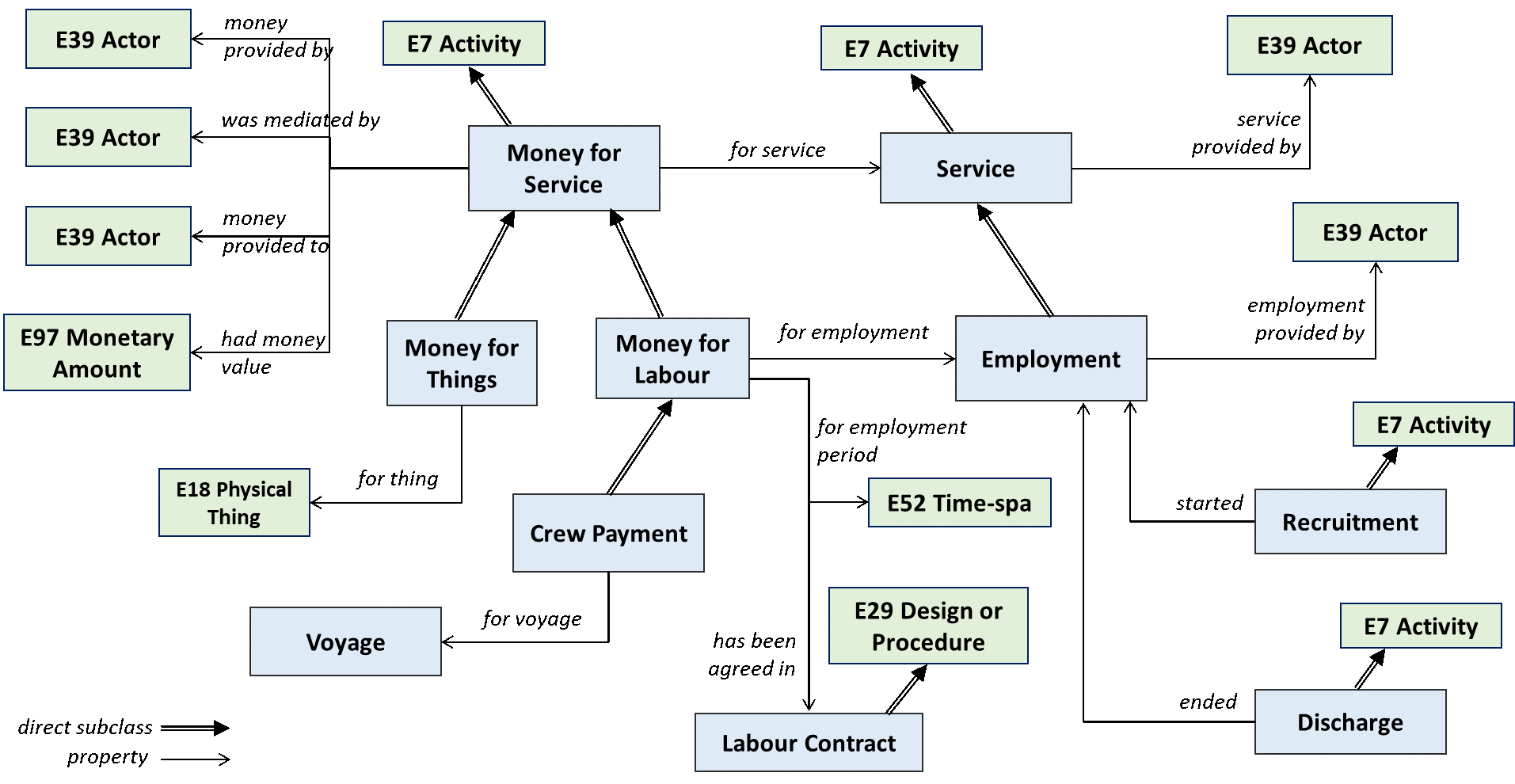}}
    \vspace{-2mm}
    \caption{Modelling information about employments and payments.}
    \label{fig:model_payments}
\end{figure}

Fig.~\ref{fig:model_payments} shows how the ontology allows describing information about \textit{employments and payments}. {\tt Money for Service} (subclass of {\tt E7 Activity}) is given to an {\tt E39 Actor} for a particular {\tt Service} (subclass of {\tt E7 Activity}).\footnote{We use the term \sq{money} instead of \sq{payment}, because we want to indicate that there was a money transaction, e.g. using lira, franc, etc. (in older times, a payment could be conducted without the use of money, e.g. using things).} The class {\tt Money for Service} has two specialisations (subclasses): {\tt Money for Things} and {\tt Money for Labour}, while the class {\tt Employment} is a specialisation of the class {\tt Service}. A {\tt Crew Payment} concerns a particular {\tt Voyage} and is a specialisation of {\tt Money for Labour}. In this context, a {\tt Labour Contract} (subclass of {\tt E29 Design or Procedure}) specifies the conditions of {\tt Money for Labour}. An {\tt Employment} starts with a {\tt Recruitment} (subclass of {\tt E7 Activity}) and ends with a {\tt Discharge} (subclass of {\tt E7 Activity}).

\begin{figure}
    \centering
    \fbox{\includegraphics[width=15cm]{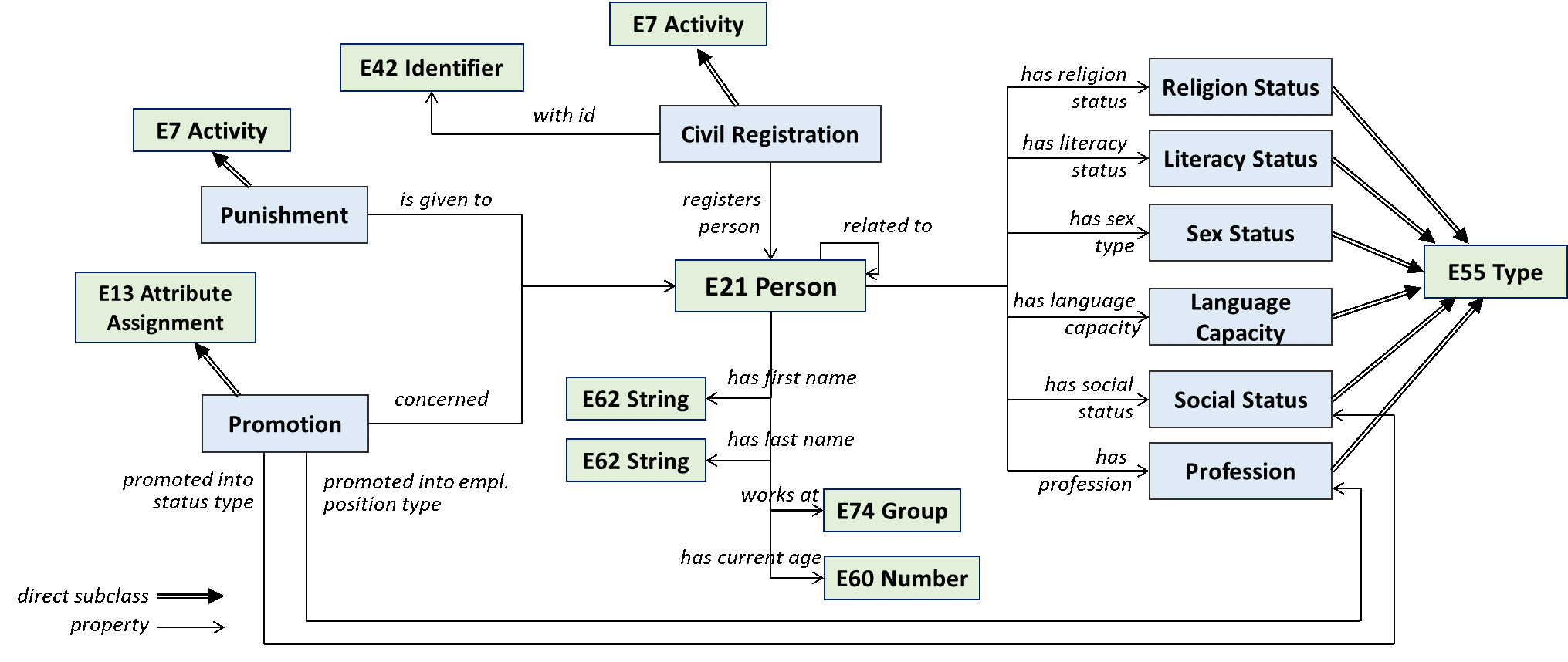}}
    \vspace{-2mm}
    \caption{Modelling information about persons.}
    \label{fig:model_persons}
\end{figure}

Fig.~\ref{fig:model_persons} shows how information about \textit{persons} (seagoing people, such as captains, crew members, students, etc.) is modelled in the ontology. A person ({\tt E21 Person}) is registered through a {\tt Civil Registration} activity and receives an identifier ({\tt E42 Identifier}). A person has a first name and last name ({\tt E62 String}), works at an organisation or company ({\tt E74 Group}), has an age ({\tt E60 Number}) at a specific time (the time of the information recording), as well as a set of other properties, in particular a {\tt Religion Status}, a {\tt Literacy Status}, a {\tt Sex Status}, a {\tt Language Capacity}, a {\tt Social Status}, and a {\tt Profession} (all subclasses of {\tt E55 Type}). The use of {\tt E55 Type} as superclass of these properties/qualities (instead of modeling them as temporal entities) is a good solution when the sources (such as a civil register or a census document) do not provide enough temporal information to infer/observe the corresponding event (this is exactly the case with the archival sources of the SeaLiT project). In addition, a {\tt Punishment} (subclass of {\tt E7 Activity}) or {\tt Promotion} (subclass of {\tt E13 Attribute Assignment}) can be given to a person. A {\tt Promotion} is related either to a {\tt Social Status} promotion or to a job/career ({\tt Profession}) promotion.

\begin{figure}[t]
    \centering
    \fbox{\includegraphics[width=12.5cm]{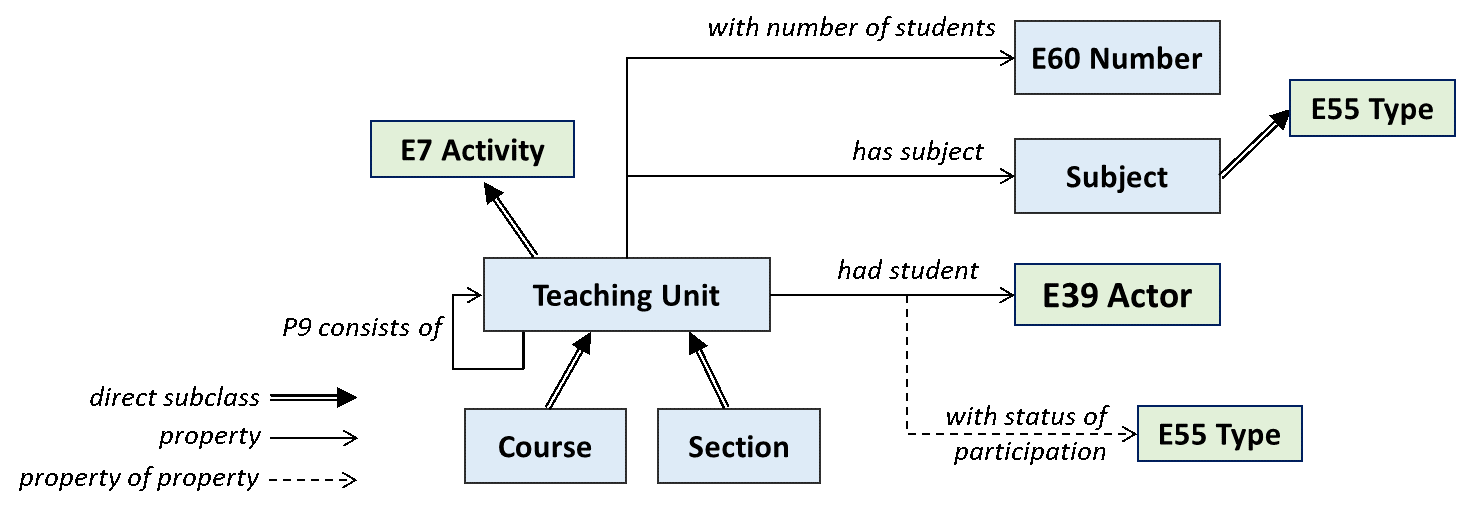}}
    \vspace{-2mm}
    \caption{Modelling information about teaching activities.}
    \label{fig:model_teaching}
\end{figure}

Finally, Fig.~\ref{fig:model_teaching} shows how the ontology allows describing information about teaching activities related to seafaring. A {\tt Teaching Unit} is an activity that can be specialised to {\tt Course} or {\tt Section}. It is connected to a {\tt Subject} (subclass of {\tt E55 Type}), the students ({\tt E39 Actor}) who participated in the teaching unit, the number of participating students ({\tt E60 Number}), as well as one or more other teaching units through the CIDOC-CRM property {\em \sq{P9 consists of}}. The latter allows, in particular, describing the information that a course consists of sections. 

\subsection{Ontology Evolution Example}
\label{subsec:evolution}

The ontology development process lasted more than two years, including a large number of intermediate versions, before releasing the first \textit{stable} version (1.0). 
In particular, the ontology elements (classes and properties) were revised several times based on 
(a) new evidence coming from newly-considered archival sources, and 
(b) new requirements (information needs) by the domain experts (maritime historians).
Such new evidence and requirements required either the definition of new elements, such as the creation of a new class or property, or the revision of an existing set of elements that concern a part of the model.

\begin{figure}[t]
    \centering
    \fbox{\includegraphics[width=16.0cm]{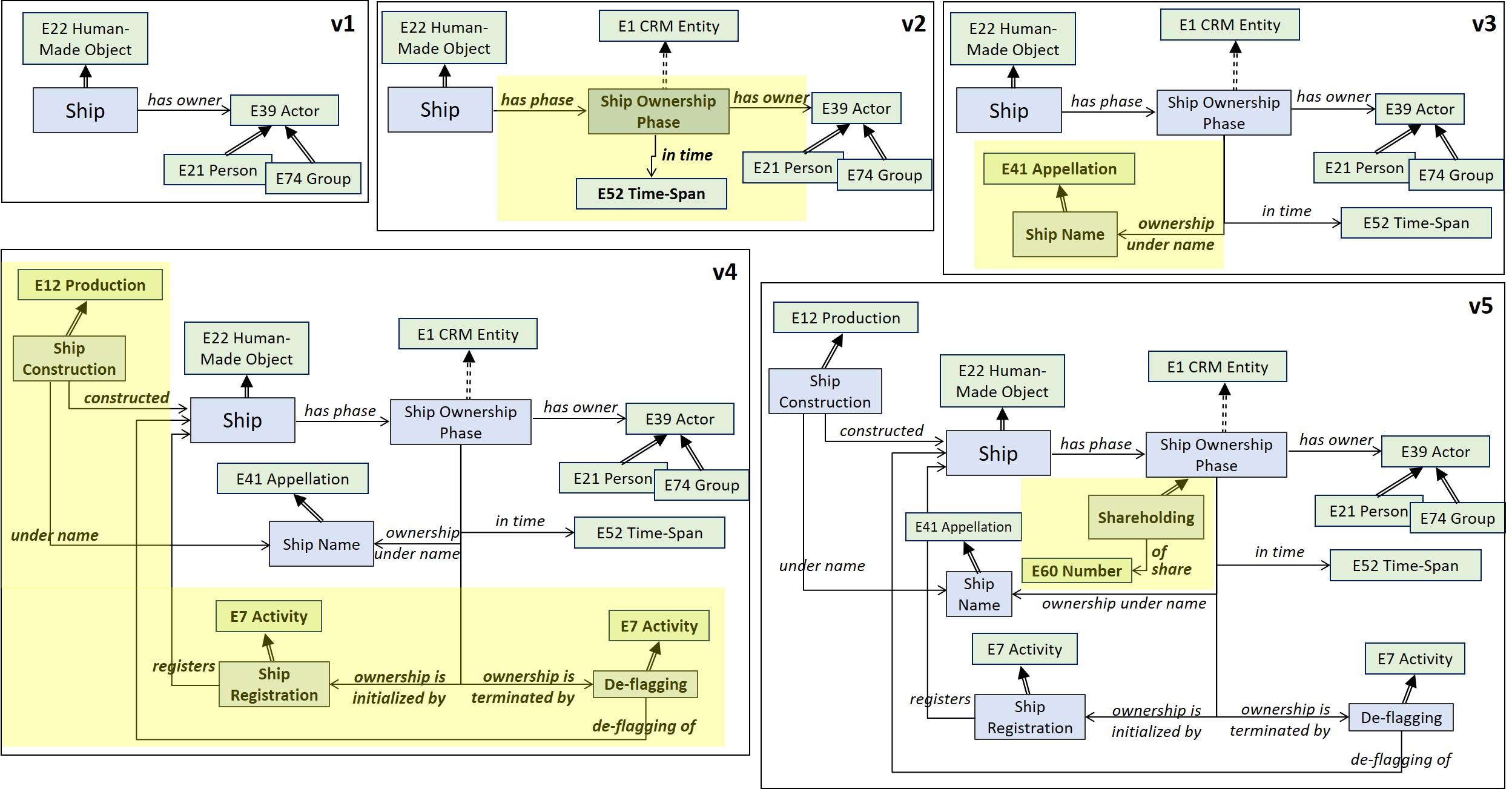}}
    \vspace{-5mm}
    \caption{Ontology evolution example for modeling ship ownership information.}
    \label{fig:evolution}
\end{figure}

Fig.~\ref{fig:evolution} shows how the part of the ontology that concerns \textit{ship ownership} was revised several times during the ontology development process. 

A first requirement provided by the historians was the ability to find all ships per owner. The analysed archival material (\textit{crew lists}) only provided the name of the owner, where the value was either the name of a person or the name of a company. Based on this evidence, the property {\em \sq{has owner}} was created  connecting an instance of {\tt Ship} with the an instance of the CIDOC-CRM class {\tt E39 Actor} (v1 in Fig.~\ref{fig:evolution}). 

Another source (\textit{naval ship register lists}) provided information about ships' previous owners, while a new requirement was the ability to find the number of first owners per ship during a period of time. Based on this, as well as on the fact that the binary relationship \textit{has owner} implies/hides a temporal entity, we defined the class {\tt Ship Ownership Phase}, the property {\em \sq{has phase}} for connecting a ship to a ship ownership phase, the property {\em \sq{in time}} for connecting the ownership phase to a {\tt E52 Time-Span}, while the property {\em \sq{has owner}} was revised for connecting the ship ownership phase with an  {\tt E39 Actor} (v2 in Fig.~\ref{fig:evolution}). 

A ship can have many names during its lifespan, while an owner can own more than one ships with the same name (as shown in \textit{logbooks} and \textit{crew and displacement lists}). According to the historians, ownership usually assigns a name to a ship and a ship changes its name under a new ownership state at a specific time. 
Based on this historical knowledge, the property {\em \sq{ownership under name}} was created for enabling to link the ship ownership phase to a {\tt Ship Name} (v3 in Fig.~\ref{fig:evolution}). 

Evidence shows that ownership of a ship is a type of information that can be inferred and not directly observed. An ownership phase can be traced by the \textit{ship registration} activity that initiates it and by the \textit{de-flagging} activity that terminates it. The documentation of a ship registration in \textit{Austrian Lloyd's fleet lists}, in particular, includes information about the ship's construction place and date, which together with the name given to ship after construction constitute safe criteria to identify a ship. 
Based on this, the classes {\tt Ship Registration} (subclass of {\tt E72 Activity}), {\tt De-flagging} (subclass of {\tt E72 Activity}) and {\tt Ship Construction} (subclass of {\tt E12 Production}) were defined, together with the properties 
{\em \sq{registers}} (for linking a registration activity to a ship),
{\em \sq{ownership is initialized by}} (for linking an ownership phase to a registration activity),
{\em \sq{de-flagging of}} (for linking a de-flagging activity to a ship),
{\em \sq{ownership is terminated by}} (for linking an ownership phase to a de-flagging activity), 
{\em \sq{constructed}} (for linking a construction activity to a ship), and 
{\em \sq{under name}} (for linking a construction activity to a ship name (v4 in Fig.~\ref{fig:evolution}).

The ownership of a ship is actually a legal agreement in which an owner holds shares. For example, according to Italian sources (\textit{maritime registers}), the ownership of a ship was structured in 24 parts (\q{carati}). Sometimes only one ship owner possessed all 24 parts. However, much more frequently the 24 parts were distributed among several ship owners. 
Based on this evidence, a new class {\tt Shareholding} was created as a specialisation (subclass) of {\tt Ship Ownership Phase}, together with the property {\em \sq{of share}} for assigning the number of shares to a shareholding phase (v5 in Fig.~\ref{fig:evolution}).

In the last ontology version (see Fig.~\ref{fig:model_ship}), {\tt Ship Ownership Phase} is defined as specialisation (subclass) of the class {\tt Legal Object Relationship}, together with the class {\tt Legal Document with Temporal Validity} which comprises official documents or legal agreements that are valid for a specific time-span. 
The more general class {\tt Legal Object Relationship} represents kinds of relationships whose state and time-span are not documented and thus cannot be directly observed. We can only observe the relationship through the events that initialise or terminate the state (starting and terminating events).

\subsection{Specification, RDFS and OWL Implementation}
\label{subsec:specAndRdfs}

The specification of the ontology and its RDFS implementation are available through the Zenodo repository (DOI: {\tt 10.5281/zenodo.6797750})\footnote{\url{https://zenodo.org/record/6797750}}, under a Creative Commons Attribution 4.0 license. 
The (resolvable) namespace of the ontology pointing to the RDFS implementation is: \url{http://www.sealitproject.eu/ontology/}. 

The specification document defines the ontology classes and properties. For each class, it provides: 
i)~its superclasses, 
ii)~its subclasses (if any), 
iii)~a scope note (a textual description of the class's intension),
iv)~one or more examples of instances of this class, and
v)~its properties (if any), each one represented by its name and the range class that it links to.
For each property, the specification provides:
i)~its domain, 
ii)~its range,
iii)~its superproperties (if any),
iv)~its subproperties (if any),
v)~a scope note, 
vi)~one or more examples of instances of this property,  and
vii)~its properties (if any).
If a property has an inverse property, this is provided in parentheses next to the property name.
Scope notes are not formal modelling constructs, but are provided to help explain the intended meaning and application of a class or property. They refer to a conceptualisation common to domain experts (maritime historians) and disambiguate between different possible interpretations. 

The RDFS implementation provides the scope note of each class or property using {\em \sq{rdfs:comment}}. 
For producing the class and property URIs, the space character in the name of a class or property is replaced by the underscore character. Inverse properties are provided using {\em \sq{owl:inverseOf}}. The version of the ontology is provided through the property {\em \sq{owl:versionInfo}} and its license through the Dublin Core term {\em \sq{dc:license}}.
For the properties pointing to classes that are represented as literals in RDF (seven properties in total, pointing to the CIDOC-CRM classes {\tt E60~Number} or {\tt E62~String}), we define their range as {\tt rdfs:Literal}. 

We also provide an OWL implementation of the ontology, containing 71 object properties, 7 datatype properties and 1 symmetric property (the property \textit{\sq{related to}}).\footnote{\url{https://sealitproject.eu/ontology/SeaLiT_Ontology_v1.1.owl}}

\begin{figure}[h]
    \centering
    \fbox{\includegraphics[width=12.5cm]{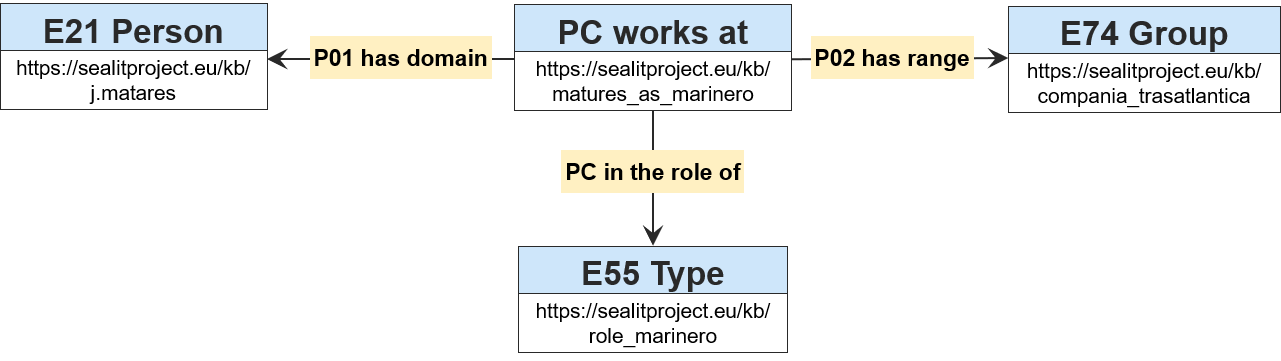}}
    \vspace{-2mm}
    \caption{Representing a property of property in RDF using a property class.}
    \label{fig:propOfProp}
\end{figure}

Since RDF does not provide a direct way to express properties of properties, we make use of \textit{property classes} (as suggested and implemented by CIDOC-CRM), as a reification method for encoding the four properties of properties defined in the SeaLiT Ontology. 
Using this method, a class is created for each property having a property. This property class can then be instantiated and used together with the properties {\em \sq{P01~has~domain}} and {\em \sq{P02~has~range}} provided by the RDFS implementation of CIDOC-CRM.\footnote{\url{https://cidoc-crm.org/rdfs/7.1.1/CIDOC_CRM_v7.1.1_PC.rdf}} 
For example, Fig.~\ref{fig:propOfProp} depicts how the property {\em \sq{in the role of}} of the property {\em \sq{works~at}} is implemented using the idea of property classes. First, the property class {\tt PC~works~at} is provided for representing the property {\em \sq{works~at}}. During data generation/instantiation, an instance of this property class is created pointing to the domain (an instance of {\tt E21~Person}) and the range (an instance of {\tt E74~Group}) of the original property {\em \sq{works~at}} using the properties {\em \sq{P01~has~domain}} and {\em \sq{P01~has~range}}, respectively. Then, we can provide the property of property {\em \sq{in the role of}} by directly linking it to the property class instance. 

\section{Application}
\label{sec:application}

\subsection{SeaLiT Knowledge Graphs}
The SeaLiT Ontology has been used in the context of the SeaLiT project (cf.~Section~\ref{subsec:sealit}) for transforming the data transcribed from a set of disparate, localised information sources of maritime history to a rich and coherent semantic network of integrated data (a \textit{knowledge graph}). The objective of this transformation is the ability to run complex questions over the integrated data, like those provided by the historians that require combining information from more than one sources.

In particular, the original archival documents are collaboratively transcribed and documented by historians in tabular form (similar to spreadsheets) using the FAST CAT system~\cite{fafalios2021FastCat}. In FAST CAT, data from different sources are transcribed as \textit{records} belonging to specific \textit{templates}. A \textit{record} organises the data and metadata of an archival document in a set of tables, while a \textit{template} represents the structure of a single data source, i.e. it defines the data entry tables.  
Currently, more than 600 records have been already created and filled in FAST CAT by historians of SeaLiT.
An example of a record for each different type of source (template) is provided in Table~\ref{tab:archSources}.

For transforming the transcribed data to RDF based on the SeaLiT Ontology, schema mappings are created for each distinct FAST CAT template. These mappings define how the data elements of the FAST CAT records (e.g. the columns of a table) are mapped to ontology classes and properties.  
To create the schema mappings and run the transformations, we make use of the X3ML mapping definition language and framework~\cite{marketakis2017x3ml}. 
The transformed data (RDF triples) are then ingested into a semantic repository (RDF triplestore) which can be accessed by external applications and services using the SPARQL language and protocol.
The ResearchSpace application (described below) operates over such a repository for supporting historians in searching and analysing quantitatively the integrated data.
The reader can refer to \cite{fafalios2021FastCat} for more information about the FAST CAT system and the data transcription, curation and transformation processes.

The generated knowledge graphs are available through the Zenodo repository (DOI: 10.5281/zenodo.6460841)\footnote{\url{https://zenodo.org/record/6460841}}, under a Creative Commons Attribution 4.0 license.
This dataset currently consists of more than 18.5M triples, providing integrated information for about 3,170 ships, 92,240 persons, 935 legal bodies, and 5,530 locations. These numbers might change in a future version since data curation, including instance matching, is still undergoing and new archival documents are transcribed in FAST CAT.

\subsection{ResearchSpace Application}

For supporting historians in exploring the SeaLiT Knowledge Graphs (and thus the integrated data), we make use of ResearchSpace~\cite{oldman2018reshaping}, an open source platform that offers a variety of functionalities, including a \textit{query building} interface that supports users in gradually building complex queries through an intuitive (user friendly) interface. The results can then be browsed, filtered, or analysed quantitatively through different visualisations, such as bar charts.
The application is accessible at: \url{http://rs.sealitproject.eu/}.

The query building interface of ResearchSpace has been configured for the case of the SeaLiT Knowledge Graphs. In particular, the following searching categories have been defined: \textit{Ship, Person, Legal Body, Crew Payment, Place, Voyage, Course, Record, Source}. By selecting a category (e.g. \textit{Ship}) the user is shown a list with its connected categories. By selecting a connected category (e.g. \textit{Place}) the user can then select a property connecting them (e.g. \textit{arrived at}) as well as an instance/value (e.g. \textit{Marseille}; thus the user is searching for ships that arrived at Marseille). Such a property actually corresponds to a path in the knowledge graph that connects instances of the selected categories.   

\begin{figure}
    \centering
        \fbox{\includegraphics[width=15.5cm]{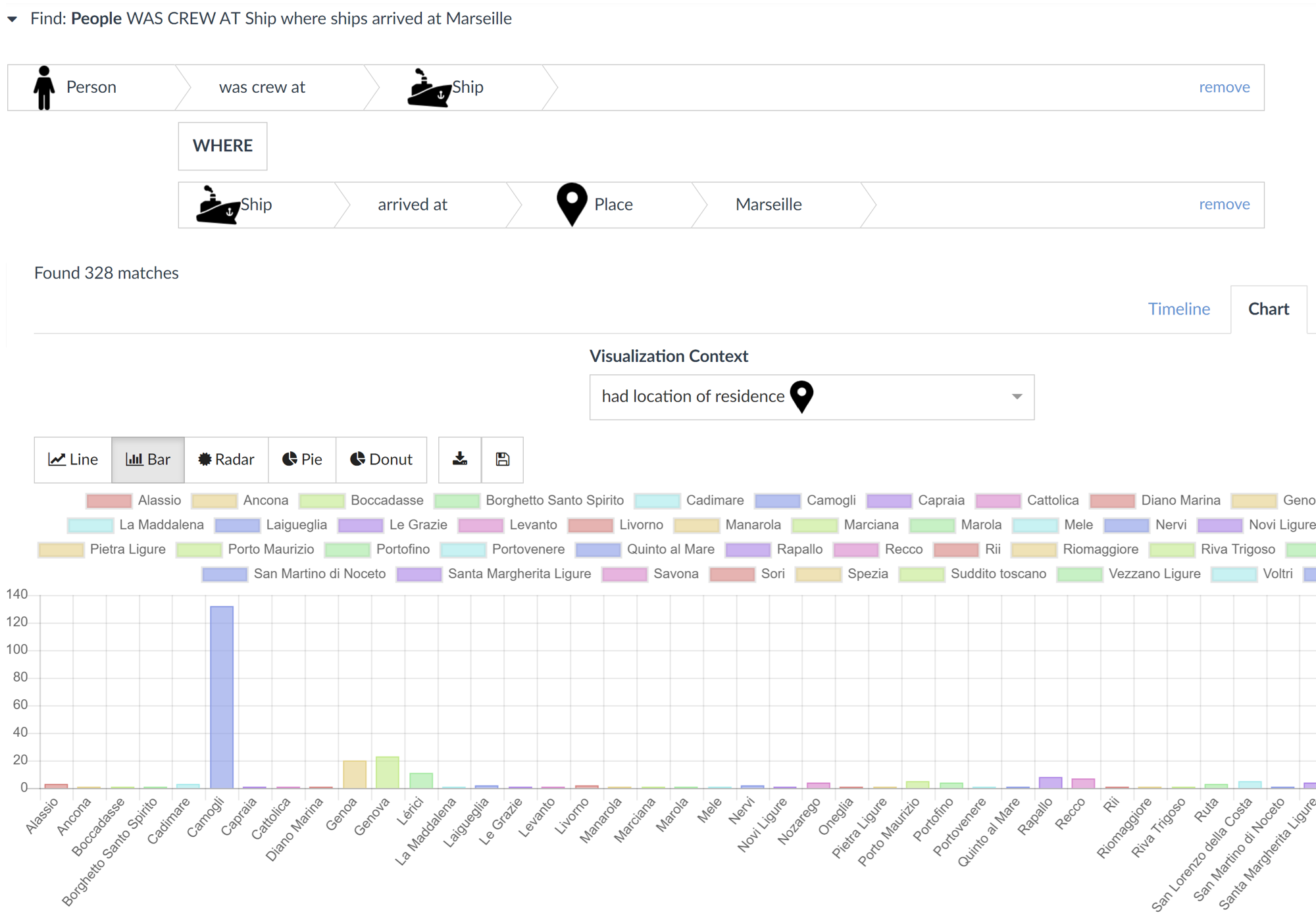}}
    \vspace{-1mm}
    \caption{Query building and visualisation of results in the ResearchSpace application.}
    \label{fig:rs}
\end{figure}

Fig.~\ref{fig:rs} shows a screen dump of the system. In this example, the user has searched for \textit{persons that were crew members at ships that arrived at Marseille,}\footnote{ResearchSpace link to the query: \url{https://tinyurl.com/2p8ky96e}} and has selected to group the persons by their \textit{residence location} and visualise the result in a bar chart.
From the bar chart we see that the majority of persons had \textit{Camogli} as their residence location.  
This query corresponds to a real information need provided by the historians of SeaLiT. 

For retrieving the results and creating the chart, ResearchSpace internally translates the user interactions to SPARQL queries that are executed over the SeaLiT Knowledge Graphs. For instance, the below SPARQL query retrieves the persons that were crew members at ships that had \textit{Marseille} as their final destination: 

\small
\begin{Verbatim}[frame=lines,numbers=left,numbersep=1pt]
 PREFIX crm: <http://www.cidoc-crm.org/cidoc-crm/>
 PREFIX sealit: <http://www.sealitproject.eu/ontology/>
 
 SELECT DISTINCT ?person WHERE {
   ?ship sealit:voyages ?voyage . 
   ?voyage sealit:finally_arriving_at <https://rs.sealitproject.eu/kb/location/Marseille> ;
           crm:P14_carried_out_by ?person }
\end{Verbatim}
\normalsize

\noindent
For grouping the persons by their residence location and showing a chart, the below SPARQL is executed for retrieving the relevant data: 

\small
\begin{Verbatim}[frame=lines,numbers=left,numbersep=1pt]
 PREFIX crm: <http://www.cidoc-crm.org/cidoc-crm/>
 PREFIX sealit: <http://www.sealitproject.eu/ontology/>
 PREFIX rdfs: <http://www.w3.org/2000/01/rdf-schema#>
 
 SELECT DISTINCT ?location ?locationName (COUNT(?person) AS ?numOfPersons) WHERE {
   ?ship sealit:voyages ?voyage  . 
   ?voyage sealit:finally_arriving_at <https://rs.sealitproject.eu/kb/location/Marseille> ;
           crm:P14_carried_out_by ?person .
   ?person crm:P74_has_current_or_former_residence ?location . 
   ?location rdfs:label ?locationName .
 } GROUP BY ?location ?locationName ORDER BY ?locationName
\end{Verbatim}
\normalsize

Such queries can also utilise the RDFS inference rules, e.g. those based on the \textit{subClassOf} and \textit{subPropertyOf} relations. An example is the use of the CIDOC CRM property \textit{\sq{P9 consists of}} for getting all voyage-related activities of a particular ship (leaving by a place, arrival at a place, passing by or through a place, loading things, unloading things), as shown in the below SPARQL query: 
\small
\begin{Verbatim}[frame=lines,numbers=left,numbersep=1pt]
 PREFIX crm: <http://www.cidoc-crm.org/cidoc-crm/>
 PREFIX sealit: <http://www.sealitproject.eu/ontology/>
 PREFIX rdfs: <http://www.w3.org/2000/01/rdf-schema#>
 
 SELECT DISTINCT ?activity ?activityName WHERE {
   <SHIP-URI> sealit:voyages ?voyage  . 
   ?voyage crm:P9_consists_of ?activity . 
   ?activity rdfs:label ?activityName }
\end{Verbatim}
\normalsize 
In this case, we exploit the fact that the property \textit{\sq{P9 consists of}} is super-property of the properties \textit{\sq{consists of leaving}}, \textit{\sq{consists of arrival}}, \textit{\sq{consists of passing}}, \textit{\sq{consists of loading}}, and \textit{\sq{consists of unloading}}.

The type of historians' research questions / information needs that can be answered (either directly or indirectly) using the ResearchSpace platform over the integrated data mainly depends on the actual archival material that is transcribed and transformed to RDF based on the SeaLiT Ontology, and less on the ontology itself.  
Specifically, the ontology was designed considering community requirements and material evidence, therefore if the data needed to answer an information need (or to find important information related to the information need) exists in the transcripts (and thus in the transformed data) then the question can be answered either fully, or partially through the retrieval of important relevant information. 
For example, in the case of SeaLiT, there are transcripts (FAST CAT records) containing tables that are not fully filled, either because some archival documents do not provide the corresponding information, or just because historians did not fill the columns during data transcription (planning to do it at a later stage). In this case, information needs that require this missing information cannot be satisfied. 
In future, if new types of information (and corresponding information needs) appear that cannot be modelled by the ontology, the ontology will be extended/revised and a new version will be released. 

With respect to incomplete information, missing entity attributes (e.g. unknown construction location for a particular ship) are in general very common in historical-archival research, but at the same time an important-to-know information for historians because they can affect the interpretation of quantitative analysis results. Our configuration of ResearchSpace considers missing information by representing it as an \sq{unknown} value, e.g. by showing an \sq{unknown} column in a bar chart.


\section{Usage and Sustainability}
\label{sec:usage}

As already stated, the ontology has been created and used in the context of the SeaLiT project for transforming data transcribed from archival documents of maritime history to a rich semantic network. The integrated data of the semantic network allows a large group of maritime historians to perform quantitative and qualitative analysis of the transcribed material (through the user-friendly interface provided by the ResearchSpace platform) and find important information relevant to their research needs.

A continuation of the relevant activities is expected after the end of the SeaLiT project through the close collaboration of the two involved institutions of the Foundation for Research and Technology - Hellas (FORTH): the Institute of Mediterranean Studies (coordinator of SeaLiT) and the Institute of Computer Science (data engineering partner in SeaLiT).
In particular, the ontology will be extended as soon as a new type of archival material needs to be transcribed and integrated into the SeaLiT Knowledge Graphs. 

The long-term sustainability of the ontology is assured through our participation in relevant communities, in particular CIDOC-CRM SIG\footnote{\url{https://www.cidoc-crm.org/sig-members}} and Data for History Consortium\footnote{\url{http://dataforhistory.org/members}}, an international consortium aiming at establishing a common method for modelling, curating and managing data in historical research. 
There is already an interest on using (and probably extending) the ontology in the context of other (ongoing) projects in the field of historical/archival research. 
In addition, the part of the model which is about employments and payments is considered for the creation of a new CIDOC-CRM family model about social transactions and bonds (there are relevant discussions on this in the CIDOC-CRM Special Interest Group; see issues 420 and 557\footnote{\url{https://cidoc-crm.org/issue_summary}}).

\section{Conclusion}
\label{sec:conclusion}

We have presented the construction and use of the SeaLiT Ontology, an extension of CIDOC-CRM for the modeling and integration of data in the field of maritime history.  
The ontology aims at facilitating a shared understanding of maritime history information, by providing a common and extensible semantic framework (a \textit{common language}) for evidence-based information integration. 
We provide the specification of the ontology, an RDFS and an OWL implementation, as well as knowledge graphs that make use of the ontology for integrating a large and diverse set of archival documents into a rich semantic network. We have also presented a real-working application (ResearchSpace deployment) that operates on top of the knowledge graphs and which supports maritime historians in exploring and analysing the integrated data through a user-friendly interface. 

In the near future, we plan to 
a) investigate possible extensions of the ontology based on new data modeling requirements,
b) improve the scope notes of classes and properties in the specification document and add more examples (and then provide a new ontology version), 
c) create and make available a JSON-LD context of the ontology for use in Web-based programming environments.

\subsection*{Acknowledgements}

This work has received funding from the European Union's Horizon 2020 research and innovation programme under 
i) the Marie Sklodowska-Curie grant agreement No 890861 (Project ReKnow), 
and ii) the European Research Council (ERC) grant agreement No 714437 (Project SeaLiT).

\bibliographystyle{ACM-Reference-Format}
\bibliography{SeaLiT_Ontology__BIB}

\appendix

\section{Class Hierarchy of SeaLiT Ontology}
\label{appendix:A}
\small
\begin{Verbatim}[frame=lines]
E1 CRM Entity
- E2 Temporal Entity
- - E4 Period
- - - E5 Event
- - - - E7 Activity 
- - - - - Voyage
- - - - - Arrival
- - - - - Leaving
- - - - - Passing
- - - - - Loading
- - - - - Unloading
- - - - - De-flagging
- - - - - Discharge
- - - - - Civil Registration
- - - - - Ship Registration
- - - - - E11 Modification
- - - - - - Ship Repair
- - - - - - E12 Production
- - - - - - - Ship Construction
- - - - - Money for Service
- - - - - - Money for Things
- - - - - - Money for Labour
- - - - - - - Crew Payment
- - - - - Teaching Unit
- - - - - - Course
- - - - - - Section
- - - - - Service
- - - - - - Employment
- - - - - E13 Attribute Assignment
- - - - - - Promotion
- - - - - Punishment
- - - - - Recruitment
- E53 Place
- - Country
- E54 Dimension
- - Horsepower
- - Duration
- - Tonnage
- E77 Persistent Item
- - E39 Actor
- - - E74 Group
- - - - Port of Registry
- - E70 Thing
- - - E71 Human-Made Thing 
- - - - E24 Physical Human-Made Thing
- - - - - E22 Human-Made Object
- - - - - - Ship
- - - - - - Ammunition
- - - - E28 Conceptual Object
- - - - - E55 Type
- - - - - - Language Capacity
- - - - - - Literacy Status
- - - - - - Navigation Type
- - - - - - Profession
- - - - - - Religion Status
- - - - - - Sex Type
- - - - - - Social Status
- - - - - - Subject
- - - E72 Legal Object
- - - - E90 Symbolic Object
- - - - - E41 Appellation
- - - - - - Ship Name
- - - - - - E42 Identifier
- - - - - - - Ship ID
- - - - - E73 Information Object
- - - - - - E29 Design or Procedure
- - - - - - - Labour Contract
- Legal Object Relationship
- - Ship Ownership Phase
- - - Shareholding
- - Legal Document with Temporal Validity
\end{Verbatim}
\normalsize

\section{Property Hierarchy of SeaLiT Ontology}
\label{appendix:B}
\small
\begin{Verbatim}[frame=lines,commandchars=\\\{\}]
\textbf{\underline{PROPERTY}}                                           \textbf{\underline{DOMAIN}}                       \textbf{\underline{RANGE}}
P1 is identified by (identifies)                   E1 CRM Entity                E41 Appellation
- has ship ID (ship ID identifies)                 Ship                         Ship ID
P2 has type (is type of)                           E1 CRM Entity                E55 Type
- has navigation type (is navigation type of)      Ship                         Navigation Type
- has language capacity (is language capacity of)  E21 Person                   Language Capacity
- has literacy status (is literacy status of)      E21 Person                   Literacy Status
- has social status (is social status of)          E21 Person                   Social Status
- has sex type (is sex type of)                    E21 Person                   Sex Type
- has profession (profession of)                   E21 Person                   Profession
- has religion status (is religion status of)      E21 Person                   Religion Status
- has subject (is subject of)                      Teaching Unit                Subject
P9 consists of (forms part of)                     E4 Period                    E4 Period
- consists of leaving (leaving is part of)         Voyage                       Leaving
- consists of arrival (arrival is part of)         Voyage                       Arrival
- consists of passing (passing is part of)         Voyage                       Passing
- consists of loading (loading is part of)         Voyage                       Loading
- consists of unloading (unloading is part of)     Voyage                       Unloading
P12 occurred in the presence of (was present at)   E63 Beginning of Existence   E77 Persistent Item
- P92 brought into existence (was brought...)      E63 Beginning of Existence   E77 Persistent Item
- - P108 has produced (was produced by)            E12 Production               E24 Physical Human-Made Thing
- - - constructed (was constructed by)             Ship Construction            Ship
- P11 had participant (participated in)            E5 Event                     E39 Actor
- - P14 carried out by (performed)                 E7 Activity                  E39 Actor
- - - navigated by captain (navigated)             Voyage                       E39 Actor
- - - registered by (is responsible for...)        Ship Registration            Port of Registry
- - - money provided by (provided money)           Money for Service            E39 Actor
- - - was mediated by (was mediator of)            Money for Service            E39 Actor
- - - money provided to (received money)           Money for Service            E39 Actor
- - - service provided by (provided service)       Service                      E39 Actor
- - - - employment provided by (provided ...)      Employment                   E39 Actor
- - had student (student in)                       Teaching Unit                E39 Actor
- P31 has modified (was modified by)               E11 Modification             E18 Physical Thing
- - repaired (was repaired by)                     Ship Repair                  Ship
- voyage of (voyages)                              Voyage                       Ship
P15 was influenced by (influenced)                 E7 Activity                  E1 CRM Entity
- P17 was motivated by (motivated)                 E7 Activity                  E1 CRM Entity
- - for voyage (motivated payment)                 Crew Payment                 Voyage
P43 has dimension (is dimension of)                E70 Thing                    E54 Dimension
- has tonnage (is tonnage of)                      Ship                         Tonnage
- has horsepower (is horsepower of)                Ship                         Horsepower
P46 is composed of (forms part of)                 E18 Physical Thing           E18 Physical Thing
- has ammunition (is ammunition of)                Ship                         Ammunition
P90 has value                                      E54 Dimension                E60 Number
- has duration value                               Duration                     E60 Number
P107i is current or former member of (has...)      E39 Actor                    E74 Group
- works at (is working place of)                   E21 Person                   E74 Group
P140 assigned attribute to (was attributed by)     E13 Attribute Assignment     E1 CRM Entity
- concerned (was promoted by)                      Promotion                    E21 Person
P141 assigned (was assigned by)                    E13 Attribute Assignment     E1 CRM Entity
- promoted into status type (status type was...)   Promotion                    Social Status
- promoted into employment position type (...)     Promotion                    Profession
P173 starts before or with the end of (ends...)    E2 Temporal Entity           E2 Temporal Entity
- P174 starts before the end of (ends after...)    E2 Temporal Entity           E2 Temporal Entity
- - P175 starts before or with the start of (...)  E2 Temporal Entity           E2 Temporal Entity
- - - started (started by)                         Recruitment                  Employment
- - P184 ends before or with the end of (ends...)  E2 Temporal Entity           E2 Temporal Entity
- - - ended (ended by)                             Discharge                    Employment
P191 had duration (was duration of)                E52 Time-Span                E54 Dimension
- had duration (duration of)                       E52 Time-Span                Duration
had flag of (was flag of)                          Ship                         Country
has crew number capacity                           Ship                         E60 Number
under name (named with)                            Ship Construction            Ship Name
with ship flag of (is flag of)                     Ship Registration            Country
with ship ID (ship ID of)                          Ship Registration            Ship ID
registers (is registered by)                       Ship Registration            Ship
has owner (is owner of phase)                      Ship Ownership Phase         E39 Actor
- has shareholder (participates with share)        Shareholding                 E39 Actor
is ownership phase of (has ownership phase)        Ship Ownership Phase         Ship
- is shareholding phase of (has shareholding)      Shareholding                 Ship 
ownership under name (name with ownership)         Ship Ownership Phase         Ship Name
is initialized by (initializes)                    Legal Object Relationship    E5 Event
- ownership is initialized by (initializes...)     Ship Ownership Phase         Ship Registration
is terminated by (terminates)                      Legal Object Relationship    E5 Event
- ownership is terminated by (terminates...)       Ship Ownership Phase         De-flagging
has owner (is owner of phase)                      Ship Ownership Phase         E39 Actor
- has shareholder (participates with share)        Shareholding                 E39 Actor
of share                                           Shareholding                 E60 Number
in time (is time of)                               Legal Object Relationship    E52 Time-Span
formerly or currently possesses (is formerly...)   E39 Actor                    Legal Doc. with Temp. Validity
de-flagging of (de-flagged in)                     De-flagging                  Ship
finally arriving at (is arrival place of)          Voyage                       E53 Place
starting from (is starting place of)               Voyage                       E53 Place
destination (is destination of)                    Voyage                       E53 Place
loaded (was loaded by)                             Loading                      E18 Physical Thing
unloaded (was unloaded by)                         Unloading                    E18 Physical Thing
at place (is place of arrival)                     Arrival                      E53 Place
from place (is place of leaving)                   Leaving                      E53 Place
by place (is place of passing by)                  Passing                      E53 Place
through place (is place of passing through)        Passing                      E53 Place
for service (service of)                           Money for Service            Service
- for employment (employment from)                 Money for Labour             Employment
had money value (was price of)                     Money for Service            E97 Monetary Amount
for employment period (is employment period of)    Money for Labour             E52 Time-Span
has been agreed in (is agreement for)              Money for Labour             Labour Contract
for thing (thing of)                               Money for Things             E18 Physical Thing
has first name                                     E21 Person                   E62 String
has last name                                      E21 Person                   E62 String
has current age                                    E21 Person                   E60 Number
with ID (ID of)                                    Civil Registration           E42 Identifier
registers person (person is registered by)         Civil Registration           E21 Person
is given to (was punished by)                      Punishment                   E39 Actor
related to                                         E21 Person                   E21 Person
with number of students                            Teaching Unit                E60 Number
\end{Verbatim}
\normalsize

\end{document}